\documentclass[prl,aps,twocolumn]{revtex4}
\usepackage{graphicx}
\usepackage{multirow}
\usepackage{amsmath}

\begin{document}

\title{Magnetic competition in topological kagome magnets}
\author{Minh-Tien Tran$^{1,2}$, Duong-Bo Nguyen$^2$, Hong-Son Nguyen$^2$, and Thanh-Mai Thi Tran$^3$}
\affiliation{$^1$Institute of Physics, Vietnam Academy of Science and Technology,
Hanoi, Vietnam}
\affiliation{$^2$Graduate University of Science and Technology, Vietnam Academy of Science and Technology, Hanoi, Vietnam}
\affiliation{$^3$Faculty of Physics, Hanoi National University of Education,  Hanoi, Vietnam}

\begin{abstract}
Magnetic competition in topological kagome magnets is studied by incorporating the spin-orbit coupling, the anisotropic Hund coupling and spin exchange into the kagome lattice. Using the Bogoliubov variational principle we find the stable phases at zero and finite temperatures.
At zero temperature and in the strong Ising-Hund coupling regime, a magnetic tunability from the out-of-plane ferromagnetism (FM) to the in-plane antiferromagnetism (AFM) is achieved by a universal property of the critical in-plane Hund coupling. At two-thirds filling the phase transition from the out-of-plane FM to the in-plane AFM is accompanied by a topological transition from quantum anomalous Hall (QAH) to quantum anomalous spin Hall (QASH) effect. Nearby half filling a large anomalous Hall conductance is observed at the magnetic phase transition. At finite temperature the out-of-plane FM is stable until a crossing temperature, above which the in-plane AFM is stable, but the out-of-plane FM magnetization is still finite. This suggests a coexistence of these magnetic phases in a finite temperature range.
\end{abstract}

\maketitle

The emergent phases resulting from the interplay between magnetism and nontrivial topology are the subject of intense research interest because of their intriguing properties and substantial interest for spintronics technologies.
The kagome lattice offers a versatile platform to study such phases, because with the special lattice geometry it can host peculiar states including unconventional magnetism \cite{Sachdev,Reimers}, nontrivial topology \cite{Nagaosa,Franz}, flat band \cite{Leykam}, Dirac electrons \cite{Martin}, quantum spin liquids \cite{Balents}. With the inclusion of electron correlations and spin-orbit coupling (SOC), the kagome lattice engenders a rich interplay between unconventional magnetism and nontrivial topology.
Recently, experiments observed striking effects including large anomalous Hall effect and unusual magnetic tunability in magnetic kagome materials \cite{Higo,LiuFelser,WangLei,Hasan}. In particular, the kagome magnet Co$_3$Sn$_2$S$_2$ exhibits an out-of-plane FM ground state, but at a finite temperature before reaching the paramagnetic (PM) state an in-plane AFM appears and coexists with the out-of-plane FM \cite{Hasan}. The competition between these magnetic phases is tunable through applying either an external magnetic field or hydrostatic pressure \cite{Hasan}.

\begin{figure}[b]
\begin{tabular}{ c c c c }
(a) & & & (b) \\
\vspace{-0.3cm}
\multirow{5}{*}{
\includegraphics[height=0.28\textwidth]{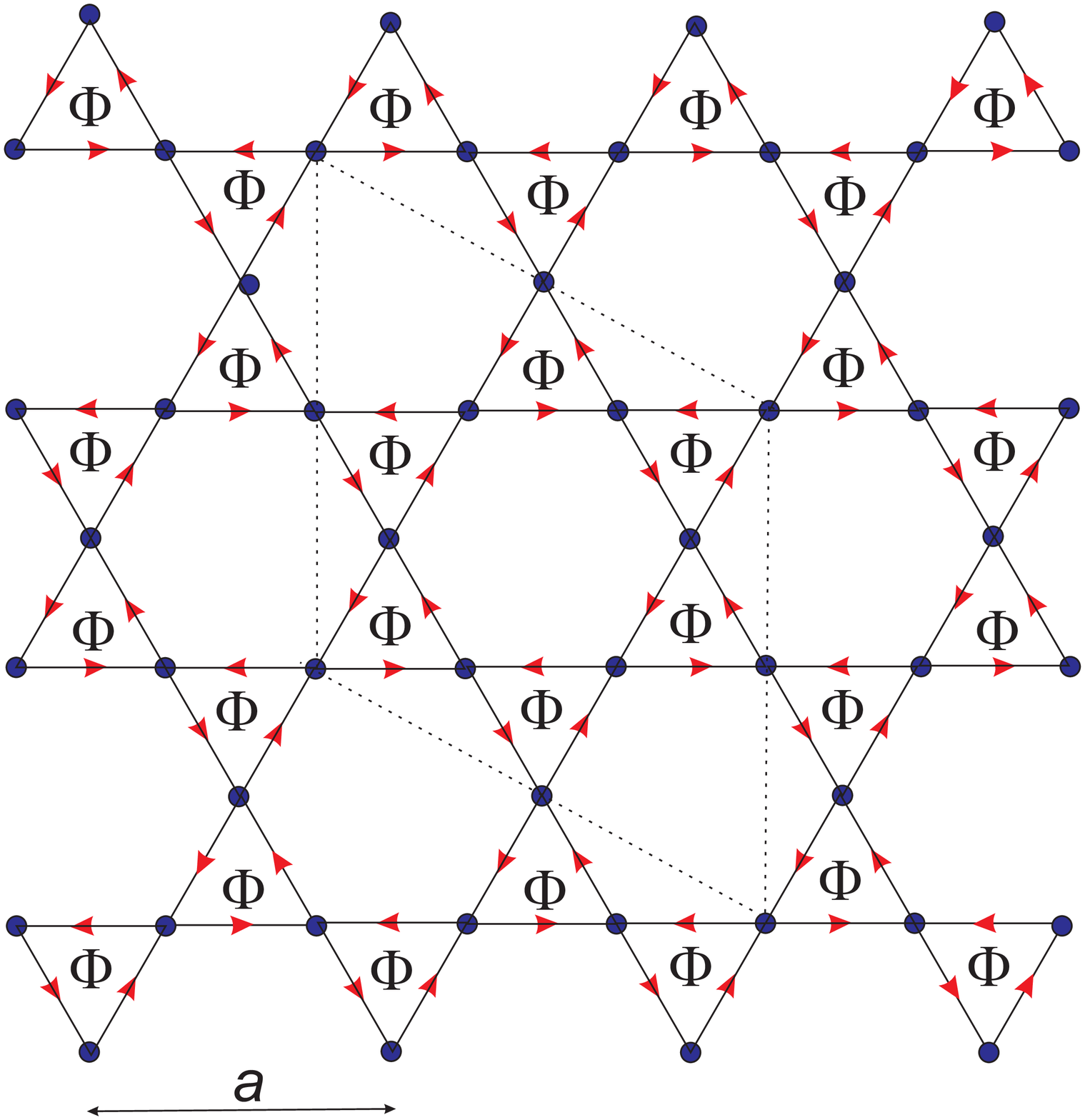}} & & & \\
 & & & \includegraphics[height=0.08\textwidth]{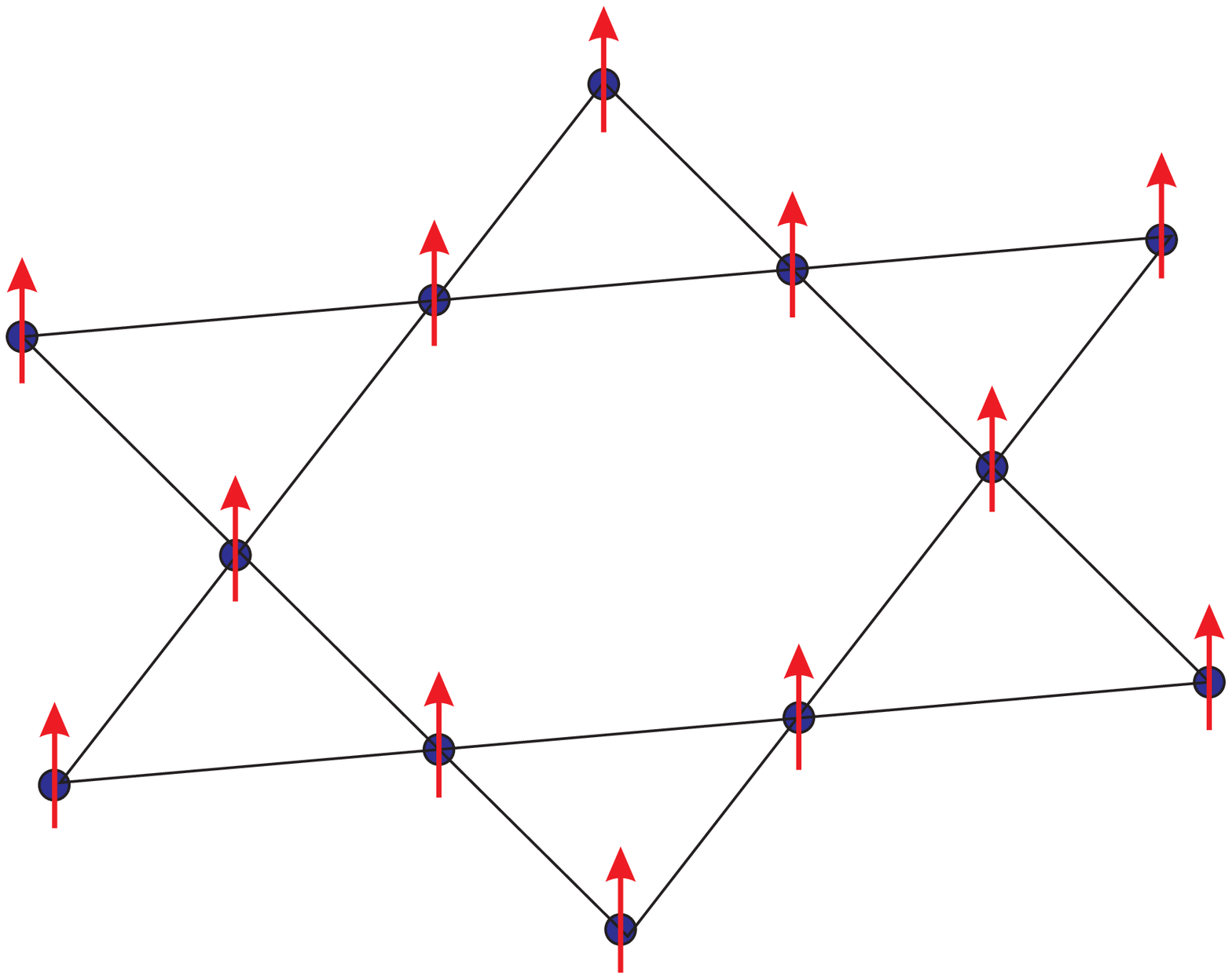} \\
 & & & (c) \\
& & & \includegraphics[height=0.08\textwidth]{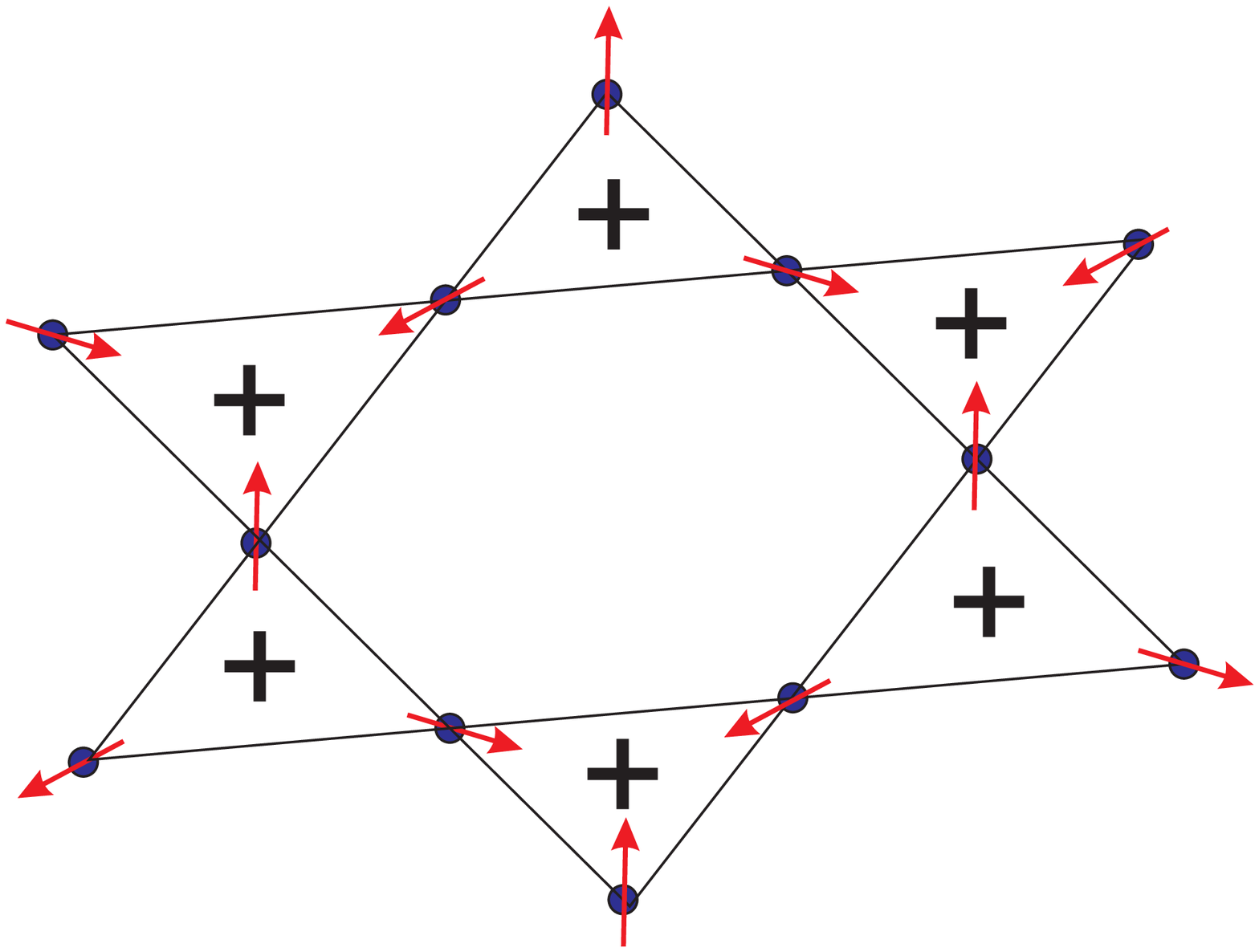} \\
& & & (d) \\
& & & \includegraphics[height=0.08\textwidth]{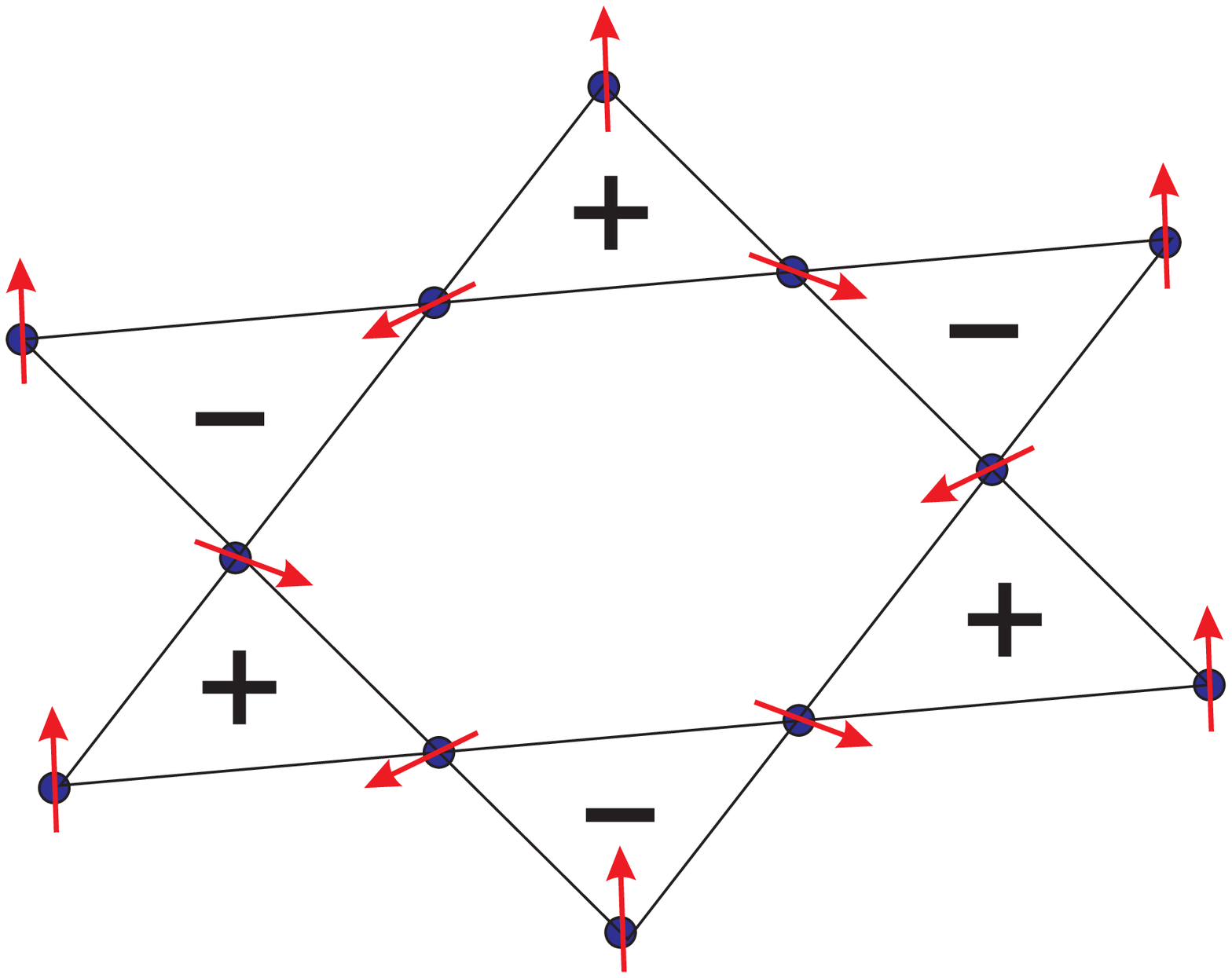}
\end{tabular}
\caption{(Color online) (a) Kagome lattice. The arrows on bonds mean the sign $\nu_{ij}=1$ of the SOC. $\Phi$ is the flux penetrating each triangle. The dotted rhombus is the $\sqrt{3} \times \sqrt{3}$ unit cell.  (b) Out-of-plane FM state. (c) \& (d) In-plane $1 \times 1$ and $\sqrt{3} \times \sqrt{3}$  AFM states, respectively. The signs $\pm$ indicate the spin chirality $\chi=\pm 1$ of each triangle.
}
\label{fig1}
\end{figure}

The present work is motivated by the striking effects observed in topological
kagome magnets, and in particular the magnetic competition between the out-of-plane FM and the in-plane AFM \cite{Higo,LiuFelser,WangLei,Hasan}.
We propose a minimal model, which can describe the observed effects.
It is generally applied to the family of topological kagome magnets, but we particularly focus on the Co$_3$Sn$_2$S$_2$ magnet. Co$_3$Sn$_2$S$_2$ has a layered crystal structure with stacked quasi-two-dimensional CoSn layers \cite{LiuFelser,WangLei,Hasan}. Magnetic Co atoms form a kagome lattice in the $xy$-plane (see Fig. \ref{fig1}).
Within our approach the electron dynamics is restricted into the two-dimensional kagome lattice.
The magnetism of Co atoms is realized through their localized spins located at
the kagome lattice sites,
reflecting strong correlations of the Co $3d$ orbitals and their Mott regime  \cite{Xu1,Xu,Hur}. Itinerant electrons come from the Sn $5p$ orbitals.  The long-range orderings of localized spins are established through their nearest-neighbor spin exchange (SE) and local Hund coupling to itinerant electrons \cite{Xu1,Xu,Hur}.
The Hamiltonian describing the kagome magnet reads
\begin{eqnarray}
H &=&-t\sum\limits_{\langle i,j \rangle, \sigma }c_{i\sigma }^{\dagger }c_{j\sigma}
-i\lambda \sum\limits_{\langle i,j \rangle ,s,s^{\prime }}
\nu _{ij}c_{is}^{\dagger }\sigma _{ss^{\prime}}^{z}c_{js^{\prime }} \nonumber \\
&&- \sum\limits_{i,\alpha,ss^{\prime }} h_{\alpha} S_{i}^{\alpha} c_{is}^{\dagger }
\sigma^{\alpha}_{ss^{\prime }} c_{is^{\prime }}
- \sum_{\langle i,j \rangle,\alpha} J_{\alpha} S^{\alpha}_i S^{\alpha}_j  ,
\label{ham}
\end{eqnarray}
where $c^{\dagger}_{i\sigma}$ ($c_{i\sigma}$) is the creation (annihilation) operator for electron with spin $\sigma$ at site $i$. $\langle i,j \rangle$ denote the nearest-neighbor lattice sites. $t$ is the hopping parameter, and $\lambda$ is the strength of the SOC. The sign $\nu_{ij}=\pm 1$ when the hopping is counterclockwise (clockwise).
$S^\alpha_i$ is the $\alpha$-component of localized spin at lattice site $i$ ($\alpha=x,y,z$), and it is renormalized that $\mathbf{S}_i^2=1$. $\sigma^\alpha$ is the Pauli matrix. $h_{\alpha}$ is the Hund coupling between the $\alpha$-component spins of itinerant and localized electrons.
$J_\alpha$ is the SE between the nearest-neighbor localized spins.
In general, we consider the case where the Hund couping and the SE are isotropic in plane $h_{x}=h_{y} \equiv h_{xy}$,   $J_{x}=J_{y} \equiv - J_{xy}$, but anisotropic out of plane. Note that the SE in plane $J_{xy}$ and out of plane $J_z$ have the opposite signs. The model in the Ising-Hund coupling limit $h_{xy}=0$ was previously proposed \cite{Hur}.

The tight-binding part of Hamiltonian in Eq. (\ref{ham}) can be rewritten as
\begin{equation}
H_{QSH} = - \sum_{\langle i,j \rangle, \sigma } t_{ij\sigma} c_{i\sigma }^{\dagger }c_{j\sigma} ,
\label{qah}
\end{equation}
where $t_{ij\sigma}= t+ i \nu_{ij} \sigma \lambda = r \exp (\pm i\Phi/3)$ with $r=\sqrt{t^2+\lambda^2}$, and $\Phi= 3 \arg(t+i\lambda)$.
In the following we use $r=1$ as the energy unit.
$\Phi$ is the flux penetrating each triangle of the kagome lattice (see Fig. \ref{fig1}a).
It ($\Phi \neq 0,\pi$) induces topologically nontrivial band structure \cite{Nagaosa,Franz}.
When the Hund coupling is included, its interplay with the SOC can emerge topological magnetic phases \cite{Tien,Tien1}.
Hamiltonian in Eq. (\ref{qah}) is just the spin version of the QAH model, which is obtained from the double exchange model in the strong Hund coupling limit \cite{Nagaosa}.
The Hall conductivity $C_\sigma$ of electrons with spin $\sigma$ in unit $e^2/h$ can be calculated
by the Kubo formula
\begin{equation}
C_\sigma \!\!=\!\! \frac{1}{N} \!\!\sum_{\mathbf{k},a,b} \!\! \frac{\Im[\langle \mathbf{k} a |j_{x} | \mathbf{k} b\rangle
\langle \mathbf{k} b |j_{y} | \mathbf{k} a\rangle
]}{(E_{\mathbf{k}a}-E_{\mathbf{k}b})^2} (f(E_{\mathbf{k}a})-f(E_{\mathbf{k}b})),
\!\!\! \label{Kubo}
\end{equation}
where $j_{\alpha}$ is the current operator in $\alpha$-direction, $|\mathbf{k} a\rangle$ and $E_{\mathbf{k}a}$ are the normalized eigenstate and eigenvalue of the Bloch Hamiltonian of electrons with spin $\sigma$, $f(x)$ is the Fermi-Dirac distribution function, and $N$ is the number of lattice sites. At zero temperature $T=0$, $C_\sigma$ is just the Chern number of the Brillouin-zone torus \cite{Kohmoto,TKNN}. Hamiltonian in Eq. (\ref{qah}) for each spin component has three bands separated by two gaps \cite{Nagaosa}. The insulating state occurs at fillings $n_\sigma=1/3, 2/3$, and $C_\sigma=\sigma$. At these fillings the charge Hall conductivity $\sigma_{xy}^{c} = (e^2/h) \sum_\sigma C_\sigma$ vanishes, whereas the spin Hall conductivity $\sigma_{xy}^{s} = (e^2/h) \sum_\sigma \sigma C_\sigma$ is quantized. This is exactly the quantum spin Hall (QSH) effect proposed in the $Z_2$ topological insulators \cite{KaneMele}.

The SE part of Hamiltonian in Eq. (\ref{ham}) is just the Heisenberg XXZ model \cite{Sachdev,Reimers}. For classical spins this model produces a magnetic phase transition from the out-of-plane FM to the in-plane AFM at $J_{xy}=2J_z$ \cite{Sachdev,Reimers,Hur}.
The spin configurations of these states are depicted in Fig. \ref{fig1}. In the out-of-plane FM all spins are parallel to the $z$-axis. The in-plane AFM states are defined within $1\times 1$ and $\sqrt{3} \times \sqrt{3}$ unit cells in the $xy$-plane (the lattice parameter is set $a=1$) \cite{Sachdev,Reimers}. These AFM states are characterized by 120$^{\circ}$ angle between spins at each triangle. They are distinguishable by the vector chirality of each triangle
$\boldsymbol{\chi}=[2/3\sqrt{3}](\mathbf{S}_1 \times \mathbf{S}_2 + \mathbf{S}_2 \times \mathbf{S}_3 +
\mathbf{S}_3 \times \mathbf{S}_1) \equiv \mathbf{e}_z \chi$, which is parallel to the $z$-axis ($\mathbf{e}_z$ is the unit vector of the $z$-axis). The $1\times 1$ AFM has the uniform chirality $\chi=1$, whereas the $\sqrt{3} \times \sqrt{3}$ AFM has
the staggered chirality $\chi=\pm 1$, as shown in Fig. \ref{fig1}. In the classical XXZ  model, the $1\times 1$ and $\sqrt{3} \times \sqrt{3}$ AFMs are degenerate. However, the in-plane Hund coupling can lift the degeneracy, as we will see later.

We will use variational calculations to find stable phases.
The variational principle is based on the Bogoliubov inequality, which reads
\begin{equation}
\Omega \leq \Omega_{tr} + \langle H - H_{tr} \rangle_{tr} \equiv \tilde{\Omega},
\label{bogoliubov}
\end{equation}
where $\Omega$, $\Omega_{tr}$ are the grand potentials corresponding to the studied $H$ and trial $H_{tr}$ Hamiltonian, respectively \cite{Feynman,Callen}. The thermodynamical average is taken over the ensemble defined by the trial Hamiltonian. Minimizing $\tilde{\Omega}$ one would find the stable phases of the studied system.

\begin{figure}[t]
\begin{center}
\includegraphics[width=0.45\textwidth]{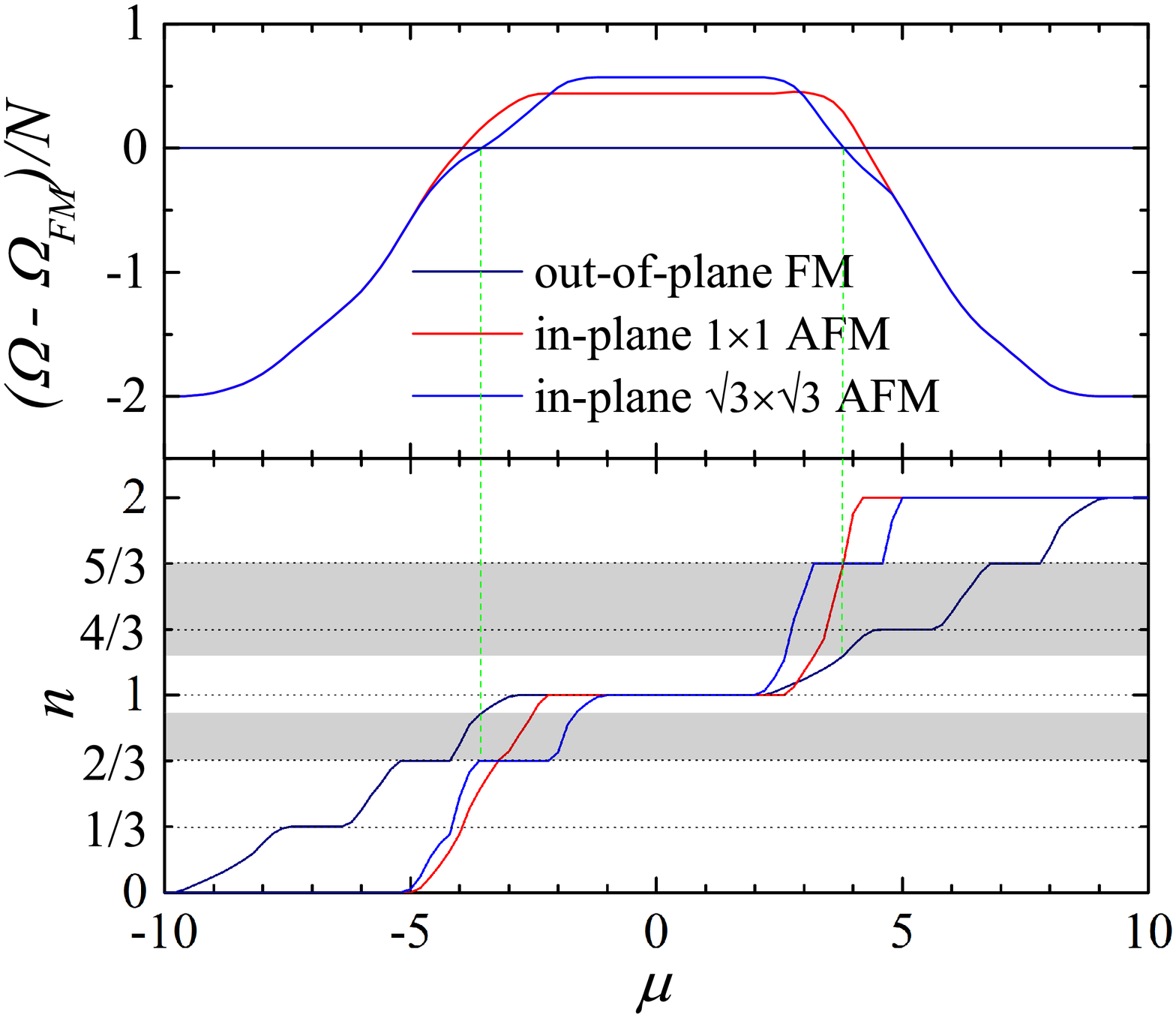}
\end{center}
\caption{(Color online) The grand potential $\Omega-\Omega_{FM}$ measured from that of the out-of-plane FM state and the electron filling $n$
at temperature $T=0$. The trial variational states are the out-of-plane FM, the in-plane $1 \times 1$ and $\sqrt{3} \times \sqrt{3}$ AFM.
The vertical dotted lines indicate the phase crossing points between the out-of-plane FM and the in-plane $\sqrt{3} \times \sqrt{3}$ AFM.
The grey shaded areas indicate the phase separation, where the electron filling is uncertain.
Model parameters $h_{z}=6$, $h_{xy}=3$, $J_{z}=1$, $J_{xy}=4$, $\Phi=\pi/3$.}
\label{fig2}
\end{figure}

\begin{figure}[t]
\begin{center}
\includegraphics[width=0.4\textwidth]{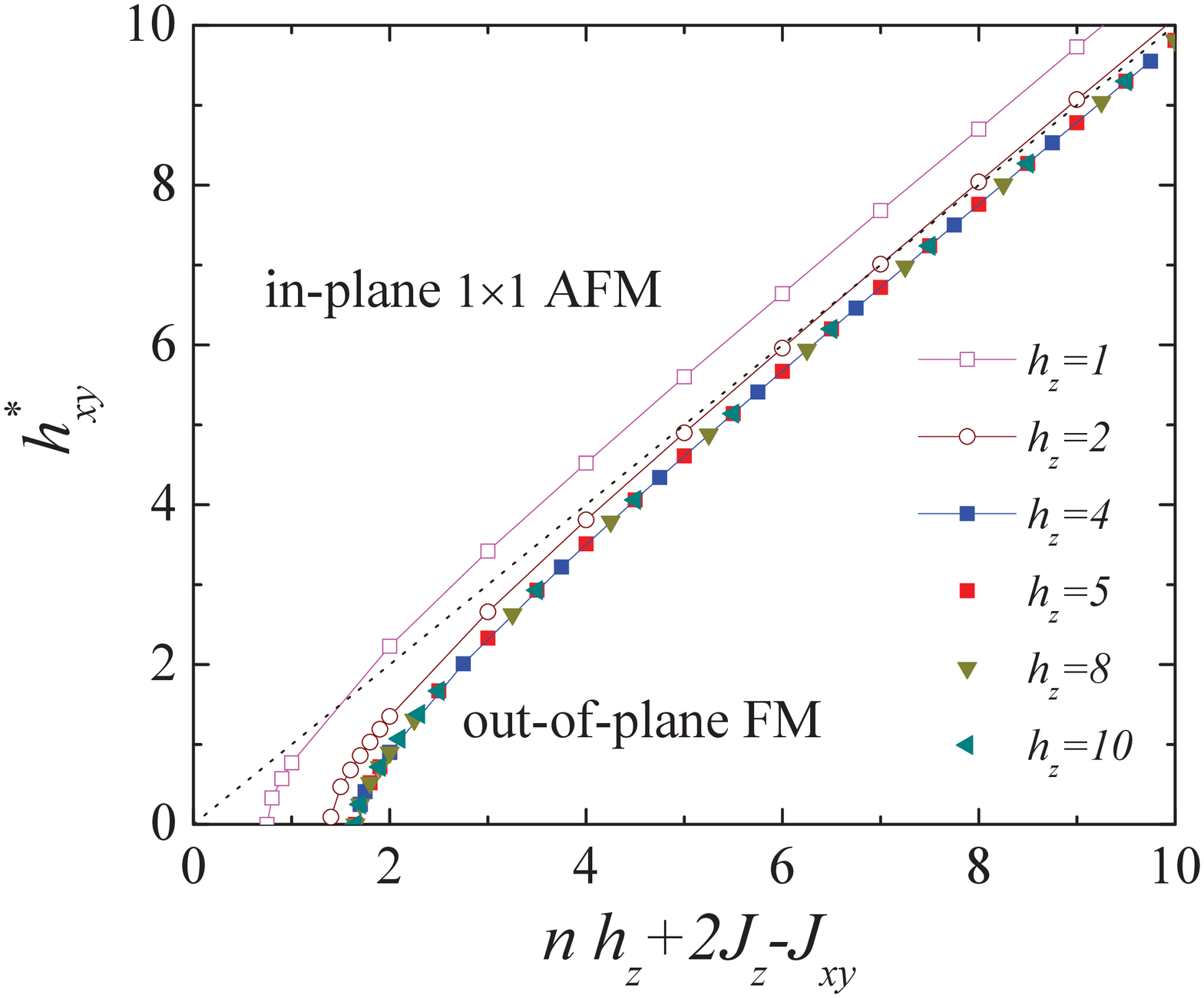}
\end{center}
\caption{(Color online) The critical line $h_{xy}^{*}$ via $(n h_z + 2 J_{z}- J_{xy})$ at half filling $n=1$ and flux $\Phi=\pi/3$.
The ground state is the out-of-plane FM when $h_{xy} < h_{xy}^*$, and the in-plane $1 \times 1$ AFM when $h_{xy} > h_{xy}^*$.
The dotted line is the asymptotic $h^*_{xy}=h_z + 2 J_{z}- J_{xy}$.
}
\label{fig3}
\end{figure}

At zero temperature $T=0$, $\tilde{\Omega}= E - \mu n N$, where $\mu$ is the chemical potential, $E$ and $n$ are the ground-state energy and the electron filling of the trial state.
In calculating the ground state energy and the electron filling,
the two-dimensional tetrahedron method is used to calculate the integration over the Brillouin zone \cite{Seki}.
We consider different trial states and find the state with lowest grand potential. It turns out that  the out-of-plane FM and the in-plane AFM are the candidates. An example of the grand potentials of trial states is shown in Fig. \ref{fig2}.
One can notice that the out-of-plane $h_z$ and $J_z$ only affect the grand potential of the out-of-plane FM, while the in-plane $h_{xy}$ and $J_{xy}$ only affect the one of the in-plane AFM. The Ising-Hund coupling $h_z$ polarizes itinerant electrons and shifts the energy levels of electrons with opposite spins in the opposite directions \cite{Nham}. A strong Ising-Hund coupling could separate the energy bands of electrons with spin up and down. The band gap appearing at half filling $n=1$ indicates this separation of the FM state. The other band gaps of the FM at fillings $n=1/3, 2/3$ or $n=4/3, 5/3$ are just similar to the ones in the non-interacting case $H_{QSH}$. However, when the Ising-Hund coupling is weak, some band gaps of the FM are closed. A similar situation is also observed for the in-plane Hund coupling and the AFM. When the  in-plane Hund coupling $h_{xy}$ is strong, it opens band gaps at fillings $n=4/3, 1, 5/3$, as shown in Fig. \ref{fig2}. However, in the weak in-plane Hund coupling regime, it can only open the band gaps at fillings $n=2/3, 4/3$, like the ones in the non-interacting case $H_{QSH}$.
We observed a magnetic competition between the out-of-plane FM and the in-plane AFM. The magnetic phase transition occurs at a critical value $h_{xy}^*$. When $h_{xy} < h_{xy}^*$, the out-of-plane FM is stable, and when $h_{xy} > h_{xy}^*$ the in-plane AFM is stable. The stable in-plane AFM is $1 \times 1$ type at half filling $n=1$ , and is $\sqrt{3} \times \sqrt{3}$ type at fillings $n=2/3, 5/3$. When $h_{xy}=0$, the in-plane $1 \times 1$ and $\sqrt{3} \times \sqrt{3}$ AFM states are degenerate. The finite in-plane Hund coupling $h_{xy}$ lifts the degeneracy.
In addition to the magnetic states, we observed also phase separations (PS) at the jumps of the electron fillings \cite{Yunoki}. They occur at the phase boundary between the phases with different symmetries such as FM and AFM. At the PS, the electron filling is uncertain and the ground state is spontaneously separated into two states with electron fillings equaled the ones at the filling jump ends.

\begin{figure}[t]
\begin{center}
\includegraphics[width=0.4\textwidth]{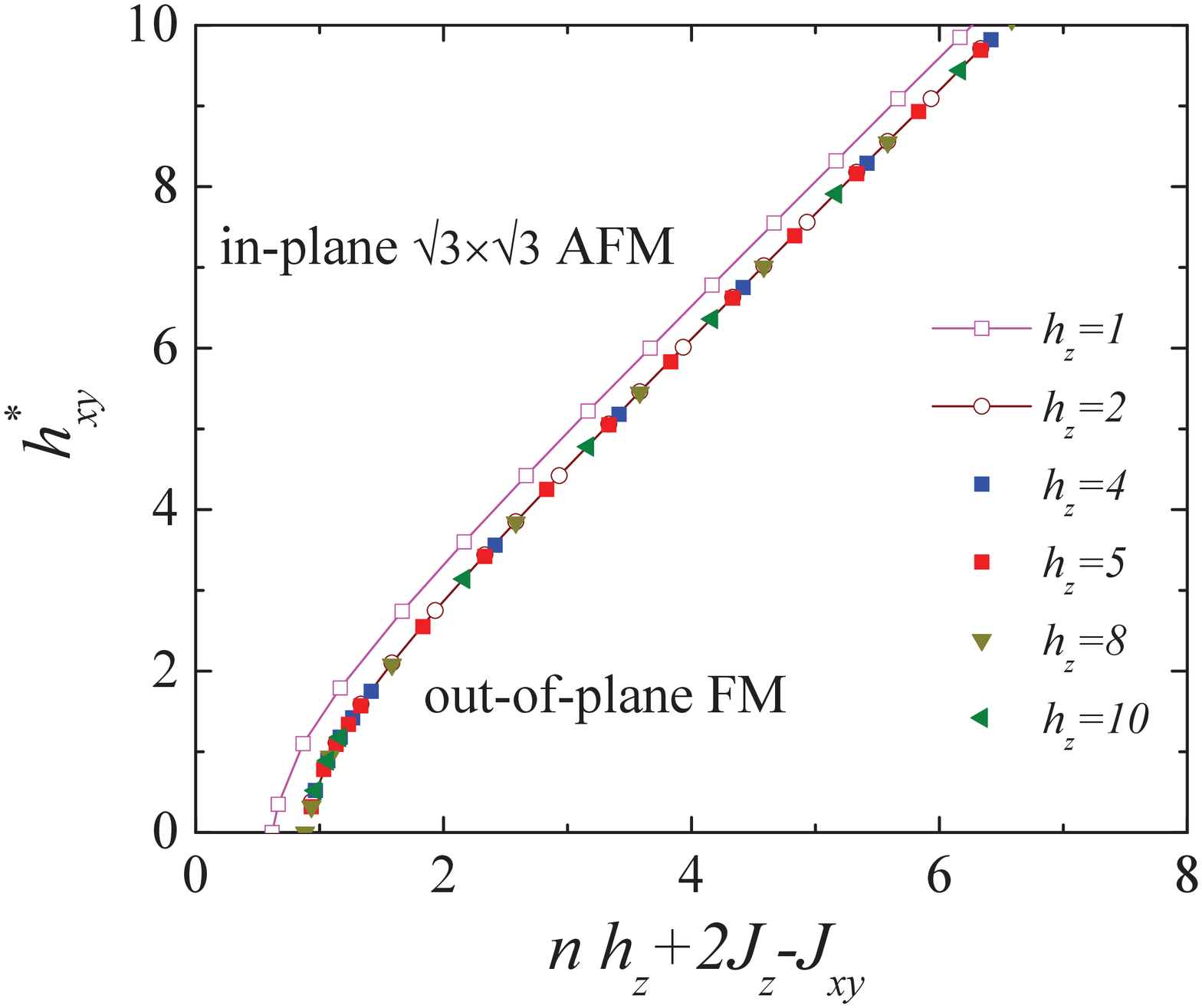}
\end{center}
\caption{(Color online) The critical line $h_{xy}^{*}$ via $(n h_z + 2 J_{z}- J_{xy})$ at filling $n=2/3$ and flux $\Phi=\pi/3$.
The ground state is the out-of-plane FM when $h_{xy} < h_{xy}^*$, and the in-plane $\sqrt{3} \times \sqrt{3}$  AFM when $h_{xy} > h_{xy}^*$. }
\label{fig4}
\end{figure}

The phase diagrams at half and two thirds fillings are summarized in Figs. \ref{fig3} and \ref{fig4}. In the regime of strong Ising-Hund coupling ($ h_{z}/n > 3$) we observed that the
critical value $h_{xy}^*$ is a universal function of
$\Delta h_z \equiv n h_{z} + 2 J_z - J_{xy}$ as shown in Figs. \ref{fig3} and \ref{fig4}.
At half filling the universal function $h_{xy}^*(\Delta h_z)$ approaches the asymptotic  $h_{xy}^* = \Delta h_{z} $ at large $\Delta h_{z}$.
The universal property of the critical $h_{xy}^*$ suggests the equivalence between the Hund coupling and the SE, as well as between their out-of-plane and in-plane components in the magnetic competition. Due to the universal property, the magnetic phase transition from the out-of-plane FM to the in-plane AFM is tunable through tuning the components of either the Hund coupling or the SE.
This also allows us to drop the SE in studying the magnetic phase transition when the Ising-Hund coupling is strong. Without the SE, the many-body local methods such as the dynamical mean field theory can safely be used \cite{Metzner}. Actually, the SE between localized spins can be generated by the Hund coupling through the Ruderman-Kittel-Kasuya-Yosida mechanism, and it is already implicitly present
in the models having the Hund coupling \cite{Ruderman,Kasuya,Yosida}.
The universal function $h_{xy}^*(\Delta h_z)$ also indicates that the kagome magnets with strong Ising-Hund coupling form a universal class, where the magnetic phase transition from out-of-plane to the in-plane magnetism does not depend on specific values of the Hund coupling and the SE as long as the value $\Delta h_z$ is fixed.
The magnetic tunability observed in Co$_3$Sn$_2$S$_2$ suggests that this kagome magnet has a strong anisotropic Hund coupling and belongs to the universal class \cite{Hasan}.

\begin{figure}[t]
\begin{center}
\includegraphics[width=0.4\textwidth]{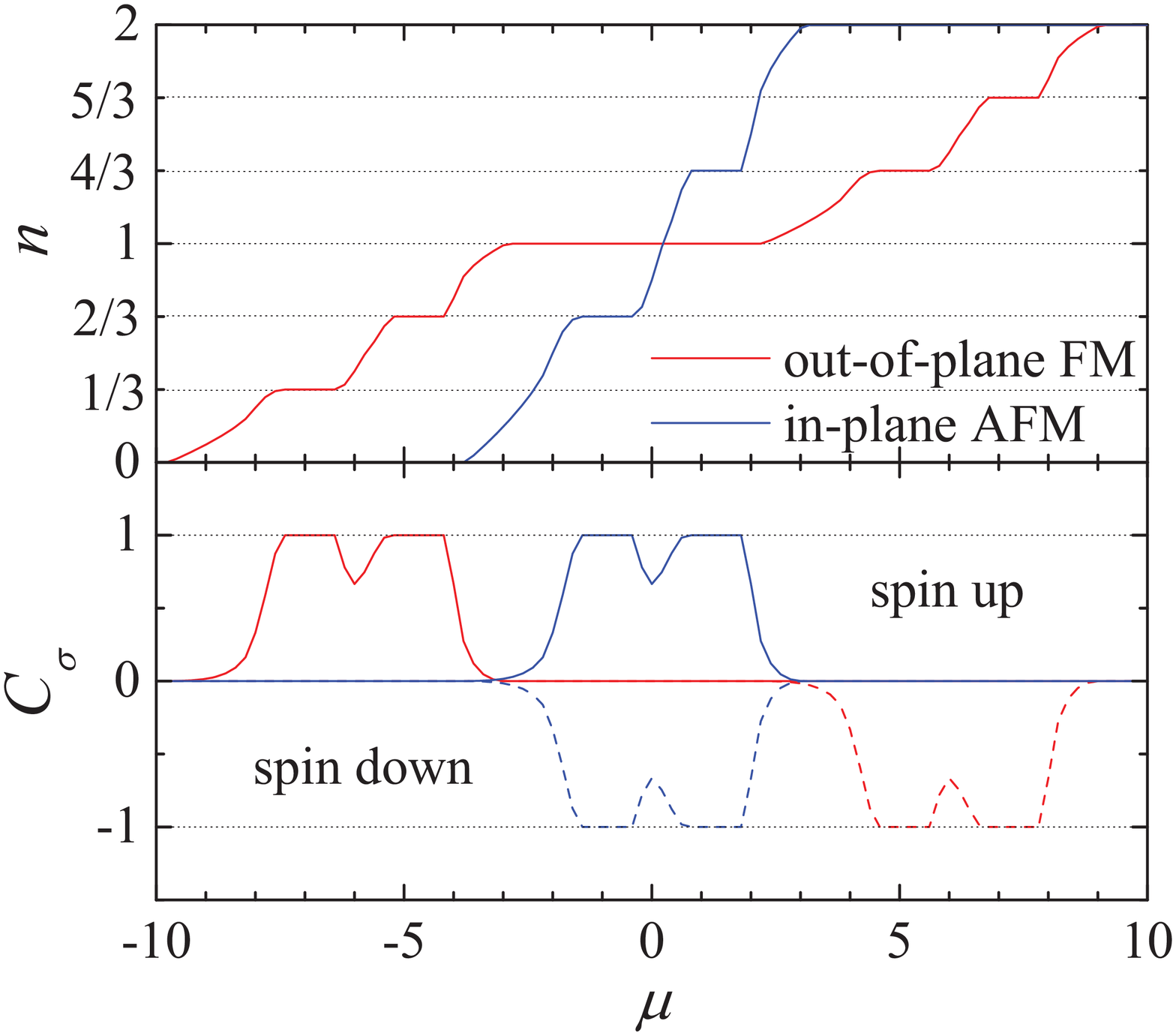}
\end{center}
\caption{(Color online) The electron filling $n$ and the spin components of the Hall conductivity $C_\sigma$ in unit $e^2/h$ via the chemical potential $\mu$ at zero temperature $T=0$ in the Ising-Hund coupling limit $h_{xy}=0$.
Model parameter $h_z=6$. }
\label{fig5}
\end{figure}

From the universal property of $h_{xy}^*(\Delta h_z)$, one can see that the magnetic phase transition still occurs in the Ising-Hund coupling limit $h_{xy}=0$, providing the in-plane spin exchange $J_{xy}$ with tunability. Thus, we can study the topological properties of the stable states in the Ising-Hund coupling limit. In this limit, the Bloch Hamiltonian of itinerant electrons in a fixed configuration of localized spins is diagonal in the spin index. Therefore,
the Hall conductivity can be separated into the spin-component $C_\sigma$, which can still be calculated
by the Kubo formula (\ref{Kubo}). Both the electron filling and the Hall conductivity are independent of the SE. However, the SE affects the ground-state energy, and it can drive the magnetic phase transition.
In Fig. \ref{fig5} we plot the electron filling and the spin components of the Hall conductivity as a function of the chemical potential at $T=0$. It shows that the out-of-plane FM has quantized $C_\sigma=\pm 1$ at fillings $n=1/3, 2/3, 4/3, 5/3$ and the in-plane AFM has quantized $C_\sigma = \sigma$ at fillings $n=2/3, 4/3$.
Therefore, the charge Hall conductivity $\sigma_{xy}^{c}$ is only quantized in the out-of-plane FM and vanishes in the in-plane AFM. However, the spin Hall conductivity $\sigma_{xy}^{s}$ is quantized in the in-plane AFM. The magnetic phase transition at two-thirds filling is also accompanied by the topological transition from QAH to QASH effect.
Nearby half filling, both the charge and spin Hall conductivities in the out-of-plane FM vanish. However, the spin Hall conductivity in the in-plane AFM is finite, although it is not quantized. It yields a large anomalous spin Hall conductance because $e^2/h a \sim 717$ $\Omega^{-1}$ cm$^{-1}$ with typical lattice parameter
$a \sim 5.4$ \AA \; \cite{LiuFelser}.
The finite value of the spin Hall conductivity nearby half filling results from an interference of two QSH conductivities at fillings $n=2/3$ and $n=4/3$. A similar finite value of the charge Hall conductivity also appears in the out-of-plane FM metals nearby fillings $n=1/3, 4/3$.

\begin{figure}[t]
\begin{center}
\includegraphics[width=0.47\textwidth]{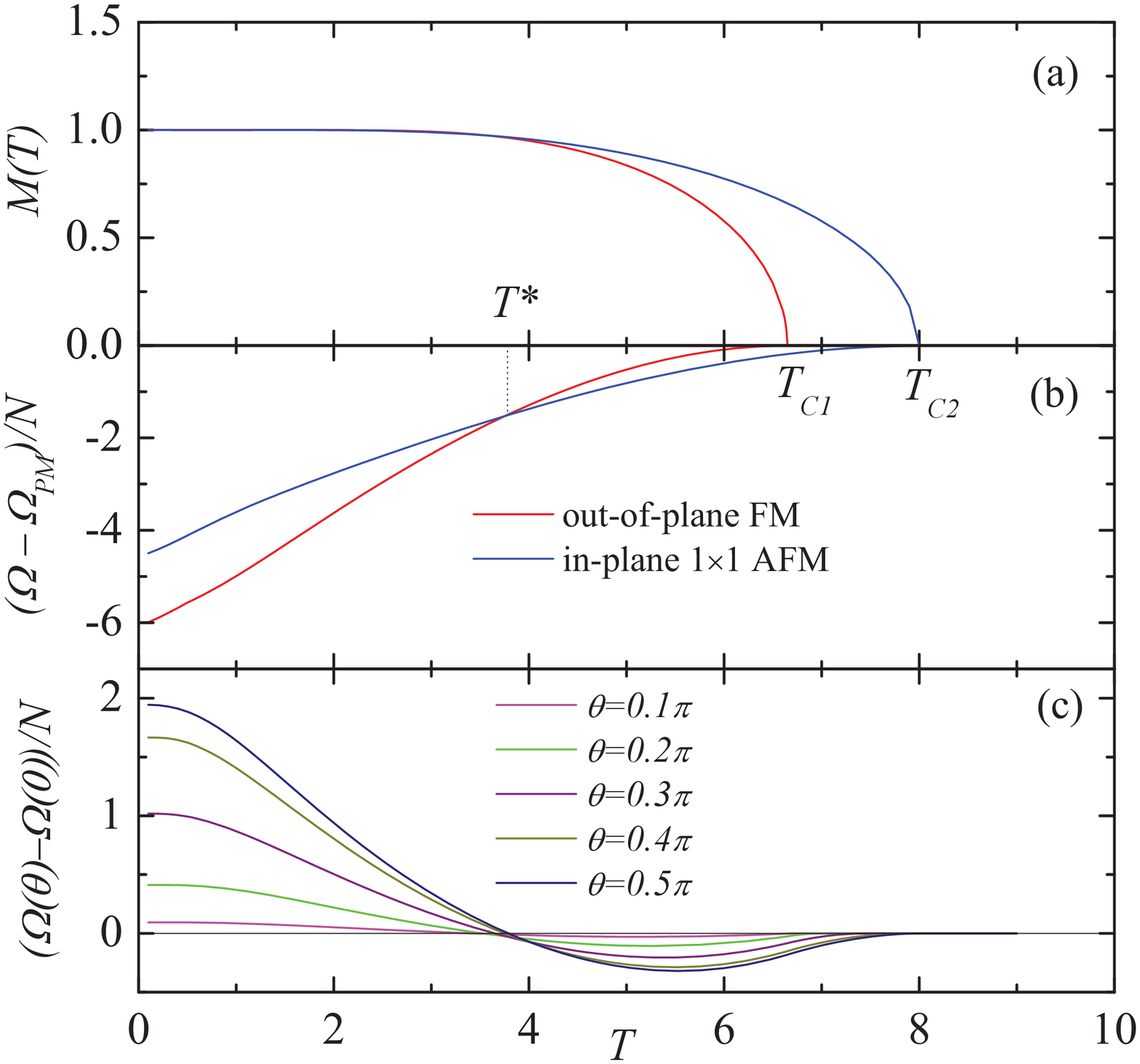}
\end{center}
\caption{(Color online) (a) The out-of-plane FM and the in-plane $1 \times 1$ AFM magnetizations $M(T)$ via temperature $T$.
(b) The grand potential $\Omega - \Omega_{PM}$ measured from that of the PM state via temperature $T$.
(c) The grand potential $\Omega(\theta) - \Omega(0)$ measured from that of the out-of-plane FM ($\theta=0$) via temperature $T$.
$T^{*}$ is the temperature below which the out-of-plane FM is stable. $T_{C1}$ and $T_{C2}$ are the critical temperatures, where the out-of-plane FM and the in-plane AFM magnetizations respectively vanish. Model parameters $h_z=6$, $h_{xy}=1$, $J_z=1$, $J_{xy}=4$, and half filling $n=1$.}
\label{fig6}
\end{figure}

At finite temperature we use the following trial ansatz
\begin{eqnarray}
\mathbf{S}_{1} &=& M(T) (-\frac{\sqrt{3}}{2} \sin\theta , - \frac{1}{2}\sin\theta, \cos\theta ) ,\\
\mathbf{S}_{2} &=& M(T) (\frac{\sqrt{3}}{2} \sin\theta , - \frac{1}{2}\sin\theta, \cos\theta ) , \\
\mathbf{S}_{3} &=& M(T) (0 , \sin\theta, \cos\theta ) ,
\end{eqnarray}
for localized spins at the triangle of the unit cell. $M(T)$ is the temperature dependence of the localized electron magnetization. $\theta$ is the angle between the spins and the $z$-axis.
The ansatz gives the out-of-plane FM when $\theta=0$, and the in-plane AFM when $\theta=\pi/2$.
When $\theta \neq 0, \pi/2$, it describes a 'umbrella' structure of spins, the projection of which in the $xy$-plane forms the in-plane AFM.
We also use the same ansatz for itinerant electron spins, reflecting the strong Hund coupling, but with different magnetization $m(T)$ \cite{Zener}. A mean field solution is obtained by minimizing the variational $\tilde{\Omega}$ in Eq. (\ref{bogoliubov}) with respect to $M(T)$ and $m(T)$. In Fig. \ref{fig6} we plot the mean field solutions for the out-of-plane FM and the in-plane AFM at half filling. It shows that the out-of-plane FM is stable below a crossing temperature $T^*$, where the grand potentials of the out-of-plane FM and the in-plane AFM are the same. However, at this crossing temperature the magnetization $M(T)$ of the out-of-plane FM does not vanish. When temperature increases, it decreases and vanishes at the critical temperature $T_{C1} > T^*$. The in-plane AFM is stable from the crossing temperature $T^*$ until the critical temperature $T_{C2} > T_{C1}$, where the PM state is reached. At the temperature range $T^* < T < T_{C1}$, the in-plane AFM is stable, but the magnetization of the out-of-plane FM is still finite. We interpret this temperature range as the region of the out-of-plane FM and in-plane AFM coexistence. The phase transition from the out-of-plane FM to the in-plane AFM occurs at the crossing point, which is infinitely degenerate, as can be seen in Fig. \ref{fig6}c. At the crossing point, any 'umbrella' phase with any $\theta$ has the same grand potential. Therefore, the magnetic phase transition is continuous, although the order parameter does not vanish.
Without such infinitely degenerate crossing point, the phase transition from out-of-plane to in-plane magnetism would abruptly occur.

In conclusion, we have studied the interplay between the SOC, the Hund coupling and the SE in the kagome lattice. It causes the magnetic competition between the out-of-plane FM and the in-plane AFM, and qualitatively describes a number of striking effects observed in the kagome magnets, including the magnetic tunability, large anomalous Hall conductance, coexistence of the out-of-plane FM and the in-plane AFM in a finite temperature range. In addition, we also observed that the magnetic phase transition from the out-of-plane FM to the in-plane AFM at two-thirds filling is accompanied by the topological transition from QAH to QASH effect. At finite temperature the magnetic phase transition is continuous although the order parameter does vanish. In the present work, quantum corrections to the mean field solution
are not considered yet. They may generate topological magnetic excitations, which may impact on the
interplay between the SOC, the Hund coupling and the SE.

This research is funded by Vietnam National Foundation for Science and Technology Development (NAFOSTED) under Grant No 103.01-2019.309.

\end{document}